\documentclass[epsfig]{aa}
\usepackage[utf8]{inputenc}
\usepackage[T1]{fontenc}
\usepackage{amsmath,amssymb}
\usepackage{graphicx}
\graphicspath{{img/}}
\usepackage{chemformula}
\usepackage{mathptmx}
\usepackage{lipsum}  
\usepackage{color}
\usepackage{epsfig}
\usepackage{natbib}
\def\aap{A\&A}

\begin{document}
\title{Hyperfine excitation of SH$^+$ by H}
\author{Fran\c{c}ois Lique\inst{1} \and Alexandre Zanchet\inst{2,3} \and Niyazi Bulut\inst{4} \and Javier R. Goicoechea\inst{2} \and Octavio Roncero\inst{2} }
\institute{
LOMC - UMR 6294, CNRS-Universit\'e du Havre, 25 rue Philippe Lebon,
BP 1123 - 76 063 Le Havre cedex, France; \email{francois.lique@univ-lehavre.fr} \label{lehavre} 
 \and Instituto de F{\'\i}sica Fundamental, CSIC, C/ Serrano, 123, 28006 Madrid, Spain \label{madrid}
 \and Departamento de Quimica Fisica, Facultad de Ciencias Quimicas, Universidad de Salamanca, 37008 Salamanca, Spain \label{salamanca}
 \and Firat University, Department of Physics, 23169 Elazi\~g, Turkey \label{elazig}
}
 
\date{Received  ; accepted }

\abstract
%context
{ SH$^+$ is a surprisingly widespread molecular ion in diffuse interstellar clouds. There, it plays an important role triggering the sulfur chemistry. In addition, SH$^+$ emission lines have been detected  at the UV-illuminated edges of dense molecular  clouds, \mbox{so-called} photo-dissociation regions (PDRs), and toward high-mass protostars. An accurate determination of the SH$^+$ abundance and of  the physical conditions prevailing in these energetic environments relies on knowing the  rate coefficients of inelastic collisions between SH$^+$ molecules and hydrogen atoms, hydrogen molecules, and electrons.}
%aims
{In this paper, we derive SH$^+$--H fine and hyperfine-resolved rate coefficients from the recent quantum calculations for the SH$^+$--H collisions, including inelastic, exchange and reactive processes.}
%approach
{The method used is based on the infinite order sudden approach.}
%results
{State-to-state rate coefficients between the first 31 fine levels and 61 hyperfine levels of SH$^+$ were obtained for temperatures ranging from 10 to 1000 K. Fine structure-resolved rate coefficients present a strong propensity rule in favour of $\Delta j = \Delta N$ transitions. The $\Delta j = \Delta F$ propensity rule is observed for the hyperfine transitions. }
%conclusions
 {The new rate coefficients will help significantly in the interpretation of SH$^+$ spectra from PDRs and UV-irradiated shocks where the abundance of hydrogen atoms with respect to hydrogen molecules can be significant.}

\authorrunning{Lique et al.}
\titlerunning{Hyperfine excitation of SH$^+$ by H}
\maketitle

\section{Introduction}

Submillimeter emission lines from the ground rotational state of  SH$^+$ were first detected toward  \mbox{W3 IRS5} high-mass star-forming region with Herschel/HIFI \citep{Benz:10}. In parallel, and using APEX telescope, \citet{menten:11} detected  
rotational absorption lines produced  by SH$^+$ in the low-density
($n_{\rm H}$\,$\lesssim$\,100\,cm$^{-3}$) diffuse  clouds 
in the line of sight  toward the strong continuum source SgrB2(M), in the Galactic Center.
Despite the very endothermic formation route of this hydride ion \citep[for a review see][]{Gerin:16}, subsequent absorption measurements of multiple lines of sight with Herschel demonstrated the ubiquitous presence of SH$^+$ in diffuse interstellar
clouds \citep{Godard:12}.

SH$^+$ rotational lines have been also detected in emission toward the Orion Bar photo-dissociation region (PDR) \citep{Nagy:13}, a strongly \mbox{UV-irradiated} surface of the Orion  molecular cloud \citep[e.g.,][]{Goicoechea:16}.
In warm and dense PDRs ($n_{\rm H}$\,$\gtrsim$\,10$^5$\,cm$^{-3}$) like the Bar, SH$^+$ forms by exothermic reactions of S$^+$ with vibrationally excited  H$_2$ \citep[with $v\,\geq\,2$, see details in][]{Agundez:10,Zanchet:13,Zanchet:19}.
High angular resolution images taken  with ALMA  shows  that SH$^+$  arises from a narrow layer at the 
edge of the PDR, the  photodissociation front that separates the atomic
from the molecular gas \citep{Goicoechea:17}. In these  PDR layers, the abundance of  hydrogen atoms is comparable to that of hydrogen molecules, that are continuously being photodissociated. Both H and H$_2$, together with electrons \citep[arising  from  the ionization of  carbon atoms, see e.g.,][]{Cuadrado:19} drive the collisional excitation of molecular rotational levels and atomic fine-structure levels.  

In addition to PDRs, the SH$^+$  line emission observed toward massive protostars likely arises from the \mbox{UV-irradiated} cavities of their molecular outflows \citep{Benz:10,Benz16}.
In these UV-irradiated shocks,  the density of hydrogen atoms can be high as well. 
All in all, the molecular abundances and physical conditions in these ambients where atomic and molecular hydrogen can have comparable abundances are not well understood. 

In the ISM, molecular abundances are derived from molecular line modeling. Assuming local thermodynamic equilibrium (LTE) conditions in the interstellar media with low densities is generally not a good approximation, as discussed by \cite{Roueff:13}. Hence, the population of molecular levels is driven by the competition between collisional and radiative processes. It is then essential to determine accurate collisional data between the involved molecules and the most abundant interstellar species, which are usually electrons and atomic and molecular hydrogen, in order to obtain reliably modeled spectra. 

The computation of collisional data for the SH$^+$ started recently. First, R-matrix calculations combined with the adiabatic-nuclei-rotation and Coulomb-Born approximations was used to compute electron-impact rotational rate coefficients and  hyperfine resolved rate coefficients were deduced using the infinite-order-sudden approximation \citep{Hamilton:18}. Then, time-independent close-coupling quantum scattering calculations are employed by \cite{Dagdigian:19} to compute hyperfine-resolved rate coefficients for (de-)excitation of SH$^+$ in collisions with both para- and ortho-H$_2$. 
 
Collisional data with atomic hydrogen are much more challenging to compute because of the possible reactive nature of the 
SH$^+$--H collisional system. However, recently, we overcame this difficult problem and presented quantum mechanical calculations of cross sections and rate coefficients for the rotational excitation of SH$^+$ by H, including the reactive channels \citep{Zanchet:19} using new accurate potential energy surfaces. 

Unfortunately, it was not possible to include the fine and hyperfine structure of the HS$^+(^3\Sigma^-$) molecule in the quantum dynamical calculations, whereas they are resolved in the astronomical observations leading the new set of data difficult to use in astrophysical applications. 

The aim of this work is to use the quantum state-to-state rate coefficients for the HS$^+(^3\Sigma^-)$--H inelastic collisions to generate a new set of fine and hyperfine resolved data that can be used in radiative transfer models. 
The paper is organized as follows : 
Sec. II provides a brief description theoretical approach. 
In Sec. III, we present the results. Concluding remarks are drawn in Sec. IV.   
 
\section{Computational methodology}

\subsection{Potential energy surfaces}

The collisions between SH$^+$($X^{3}\Sigma^{-}$) and H($^2S$) can take place on two different potential energy surfaces (PESs), the ground quartet ($^4A''$) and doublet ($^2A''$) electronic states of the H$_2$S$^+$ system.
In this work, we used the H$_2$S$^+$ quartet and doublet potential energy surfaces (PESs), that were previously
generated by \cite{Zanchet:19}.

Briefly, the state-average complete active space (SA-CASSCF) method \citep{Werner:85} was employed to calculate the first $^4A''$ together with the two first $^2A'$ and the three first $^2A''$  electronic states. The obtained state-average orbitals and multireference configurations were then used to calculate both the lowest $^4A''$
and $^2A''$ states energies with the internally contracted multireference configuration interaction method (ic-MRCI)  \citep{Werner:88} and Davidson correction \citep{Davidson:75}. 
For both sulfur and hydrogen atoms, the augmented correlation-consistent quintuple zeta (aug-cc-pV5Z) basis sets were used and all calculations were done using the MOLPRO suite of programs \citep{MOLPRO}. These energies have then been fitted using the GFIT3C procedure \citep{Aguado-Paniagua:92,Aguado-etal:98}.

Both PESs exhibit completely different topographies.
The $^4A''$ electronic state do not present any minimum out of the van der Waals wells in the asymptotic
channels and does not present any barrier to $\ch{SH+ + H -> H2 + S+}$ reaction.
This reaction is exothermic on this surface and
reactive collisions are likely to occur in competition with the inelastic collisions.

On the other hand, the $^2A''$ state presents a deep insertion HSH well and does not present any barrier neither.
For this state, in contrast with the previous case, the $\ch{SH+ + H -> H2 + S+}$ reaction is endothermic and only inelastic collisions can occur (pure or involving H exchange).

\subsection{Time independent and Wave Packet calculations}

During a collision between \ch{SH+} and \ch{H}, three processes compete:
	the inelastic (\ref{ine}), reactive (\ref{rea}) and exchange (\ref{ex}) processes:
	\begin{equation}
		\ch{SH+}(v,N) + \ch{H'} \ch{->} \ch{SH+}(v',N') + \ch{H'} 
		\label{ine}
	\end{equation}
	\begin{equation}
		\ch{SH+}(v,N) + \ch{H'} \ch{->} \ch{H'H}(v',N') + \ch{S+} 
		\label{rea}
	\end{equation}
	\begin{equation}
		\ch{SH+}(v,N) + \ch{H'} \ch{->} \ch{H'S+}(v',N') + \ch{H} 
		\label{ex}
	\end{equation}
	where $v$ and $N$ designate the vibrational and rotational levels, respectively, of the molecule (\ch{SH+} or \ch{H2}  when the reaction occurs). Only collisions with \ch{SH+} molecules in their ground vibrational state $v=0$ are considered in this work. Therefore, the vibrational quantum number $v$ will be omitted hereafter. 
	
The spin-orbit couplings between the different H$_2$S$^+$ were ignored and the collision on the ground quartet and doublet electronic states were studied separately. Because of their different topography, the dynamical calculations were treated differently on the two PESs.

The reaction dynamics on the $^4A''$ state has been studied with a time-independent treatment
based on hyperspherical coordinates. On this PES, the $\ch{SH+ + H -> H2 + S+}$ collision is a barrierless and exothermic reaction for which it has been shown \citep{Zanchet:19} that the reactivity is large ($k > 10^{-10}$ cm$^{3}$ s$^{-1}$), even at low temperatures. Hence, the competition between all the three processes (inelastic, exchange and reactivity) is taken into account rigorously. We used the \textsc{abc} reactive code of \citet{skouteris:2000} to carry out close coupling calculations of the reactive, inelastic and exchange cross sections. The cross sections were obtained following the approach described in \citet{tao:2007} and recently used to study the rotational excitation of \ch{HeH+} \citep{Desrousseaux:20} by \ch{H}. 

We computed the cross sections for the first 13 rotational levels of the \ch{SH+} molecule ($0< N <12$) for collisional energy ranging from 0 to 5000 cm$^{-1}$ and for all values of the total angular momentum $J$ leading to a non-zero contribution in the cross sections. More details about the scattering calculations can be found in \cite{Zanchet:19}.

The ground doublet ($^2A''$) electronic states exhibit a large well depth. Then, time-independent treatment is not usable and the dynamics was studied from a quantum wave-packet method using the MAD-WAVE3 program \cite{Zanchet-etal:09b}. On this electronic state, the reactive channels are largely endothermic and are not open at the collisional energies considered in this work.

The inelastic and exchange cross sections on the $^2A''$ state were calculated using the usual partial wave expansion as
\begin{eqnarray}
\sigma_{\alpha,N\rightarrow \alpha', N'}(E_k)
&=&\frac{\pi}{k^{2}} \frac{1}{2N+1}  \sum_{J=0}^{J_{max}} \sum_{\Omega,\Omega'}(2J+1)
\nonumber\\
&& \times  P^{J}_{\alpha vN \Omega\rightarrow \alpha' v'N'\Omega'}(E_k) \nonumber\\
\end{eqnarray}
where $J$ is the total angular momentum quantum number, and $\Omega, \Omega'$ are the projections
of the total angular momentum on the reactant and product body-fixed  z-axis, respectively. $\alpha=I, \alpha'=I\ {\rm or}\ E$
denotes the arrangement channels, inelastic or exchange. $k^2=2\mu_rE_k/\hbar^2$ is the square of the
wave vector for a collision energy $E_k$, and $P^{J}_{\alpha vN \Omega\rightarrow \alpha' v'N'\Omega'}(E_k)$
are the transition probabilities, $i.e.$ the square of the corresponding S-matrix elements. We computed the cross sections from the $N=0$ rotational states to the first 13 rotational levels of the \ch{SH+} molecule ($0< N' <12$) for collisional energy ranging from 0 to 5000 cm$^{-1}$ Because of the high computational cost of these simulations, they are only performed for $J$=0, 10, 15, 20, 25, 30, 40, 50, ..., 110, while for intermediate $J$ values they are interpolated using a uniform $J$-shifting approximation as recently used for the OH$+$--H and CH$^+$--H collisional systems \cite{Werfelli:15,bulut:15}. The convergence analysis and
the parameters used in the propagation for each of the two PESs used are described in detail in \cite{Zanchet:19}. 
  
For both sets of calculations, since the two rearrangement channels, inelastic and exchange, yields to the same products,
the corresponding cross sections are summed for the doublet and quartet states independently. Finally, the cross sections for each of the two electronic states are summed with the proper degeneracy factor to give the total collision 
cross sections as
\begin{eqnarray}
\sigma_{N \to N'} (E_{k}) = {2\over 3} \sigma^{S=3/2}_{N \to N'} (E_{k}) 
+ {1\over 3} \sigma^{S=1/2}_{N \to N'} (E_{k}).
\end{eqnarray}
As seen in \cite{Zanchet:19}, the magnitude of the excitation cross sections obtained on the doublet states are larger than that on the quartet states, because of the both, the non reactive character of the collision and of the deep well that favor inelastic collisions.

From the total collision cross sections $\sigma_{N \to N'} (E_{k})$, 
one can obtain the corresponding thermal rate coefficients at temperature $T$ 
by an average over the collision energy ($E_k$):
\begin{eqnarray}
\label{thermal_average}
k_{N \to N'}(T) & = & \left(\frac{8}{\pi\mu k_B^3 T^3}\right)^{\frac{1}{2}}  \nonumber\\
&  & \times  \int_{0}^{\infty} \sigma_{N \to N'}(E_k)\, E_{k}\, exp(-E_k/k_BT)\, dE_{k}
\end{eqnarray}
where $k_B$ is Boltzmann's constant and $\mu$ the reduced mass. The cross sections calculations carried out up to kinetic energy of 5000 cm$^{-1}$ allowed computing rate coefficients for temperatures ranging from 10 to 1000 K.

In all these calculations, the spin-rotation couplings of SH$^+$ have not been included, and therefore the present set of rate coefficients cannot be directly used to model interstellar SH$^+$ spectra where the fine and hyperfine structure is resolved. 

\subsection{Infinite order sudden (IOS) calculations}

In this section, we describe how the state-to-state fine and hyperfine rate coefficients for the SH$^+$--H collisional system were computed using IOS methods \citep{Goldflam:77,Faure:12} using the above $k_{0 \to N'}(T)$ 
rate coefficients as "fundamental" rate coefficients (those out of the lowest level)

For SH$^+$ in its ground electronic $^{3}\Sigma^{-}$ state, the molecular energy levels can be described in the Hund's case (b) limit\footnote{For $^{3}\Sigma^{-}$ electronic ground state molecules, the energy levels are usually described in the intermediate coupling scheme \citep{Gordy:84,lique:05}. However, the use of IOS scattering approach
 implies to use the Hund's case (b) limit.}. 
The fine structure levels are labeled by $Nj$, where $j$ is the total molecular angular momentum quantum number 
with $\vec{j}=\vec{N}+\vec{S}$. $\vec{S}$ is the electronic spin. 
For molecules in a  $^{3}\Sigma^{-}$ state,  $S=1$. 
 Hence, three kinds of levels ($j=N-1$,  $j=N$ and $j=N+1$) exist, 
except for the $N=0$ rotational level which is a single level. 

The hydrogen atom also possesses a non-zero nuclear spin ($I=1/2$). 
 The coupling between $\vec{I}$ and $\vec{j}$ results in a splitting
 of each level into two hyperfine levels (except for the $N=1,j=0$ level 
which is split into only one level).  Each hyperfine level is designated 
by a quantum number $F$ ($\vec{F}=\vec{I}+\vec{j}$) varying between $| I - j |$ and $I + j$.

Using the IOS approximation, rate
coefficients among fine structure levels can be obtained from the
$k_{0 \to L} (T)$ "fundamental" rate coefficients using the following formula
\citep[e.g. ][]{corey83}:
\begin{eqnarray} \label{REEQ}
k^{IOS}_{Nj \to N'j'} (T)  & = & 
(2N+1)(2N'+1)(2j'+1) \sum_{L}\nonumber \\
& & \left(\begin{array}{ccc}
N' & N & L \\ 
0 & 0 & 0
\end{array}\right)^{2} 
\left\{\begin{array}{ccc}
N & N' & L \\
j' & j & S 
\end{array}
\right\}^2 \nonumber \\
& & \times k_{0 \to L} (T) 
\end{eqnarray}
where $\left( \quad \right)$ and $\left\{ \quad \right\}$ are
respectively the ``3-j'' and ``6-j'' symbols.
In the usual IOS approach, $k_{0 \to L}(T)$ is calculated for each collision angle. 
Here, however, we use the $k_{0 \to L}(T)$ rate coefficients of Eq.~(\ref{thermal_average}) obtained with
a more accurate quantum method.

The hyperfine resolved rate coefficients can also be obtained from the fundamental rate coefficients as follow \cite{Faure:12}:
\begin{eqnarray} \label{REEQ2}
k^{IOS}_{NjF \to N'j'F'} (T)  & = & 
(2N+1)(2N'+1)(2j+1)(2j'+1)\nonumber \\
& & \times  (2F'+1) \sum_{L} \left(\begin{array}{ccc}
N' & N & L \\ 
0 & 0 & 0
\end{array}\right)^{2} \nonumber \\
& & \left\{\begin{array}{ccc}
N & N' & L \\
j' & j & S 
\end{array}
\right\}^2 
\left\{\begin{array}{ccc}
j & j' & L \\
F' & F & I 
\end{array}
\right\}^2 \nonumber \\
& & \times k_{0 \to L} (T) 
\end{eqnarray}

In addition, we note that the fundamental excitation rates
$k_{0\to L}(T)$ were in practice replaced by the de-excitation
fundamental rates using the detailed balance relation:
\begin{equation}
k_{0\to L}(T)=(2L+1)k_{L\to 0}(T)
\end{equation}
where
\begin{equation}
k_{L \to 0}(T) = k_{0 \to L}(T) \frac{1}{2L+1} e^{\frac{\varepsilon_L}{k_BT}}
\end{equation}
$\varepsilon_{L}$ is the energies of the rotational levels $L$. 

This procedure was indeed found to significantly improve the results at low temperatures due to important threshold effects.

The fine and hyperfine splittings of the rotational states are of a few cm$^{-1}$ and of a few 0.001 cm$^{-1}$, respectively and can be neglected compared to the collision energy at $T>30-50$ K so that the present approach is expected to be reasonably accurate for all the temperature range considered in this work.
\cite{Lique:16} have investigated the accuracy of the IOS approach in the case of OH$^+$--H collisions.
It was shown to be reasonably accurate (within a factor of 2), even at low temperature so that we can anticipate a similar accuracy for the present collisional system. 
In addition, we note that with the present approach, some fine and hyperfine rate coefficients are strictly zero. This selection rule is explained by the ``3-j'' and ``6-j'' Wigner symbols that vanish for some transitions. Using a more accurate approach, these rate coefficients will not be strictly zero but will generally be smaller than the other rate coefficients.
 
\section{Results}

Using the computational methodology described above, we have generated fine and hyperfine resolved rate coefficients for the SH$^+$--H collisional system using the doublet and quartet pure rotational rate coefficients in order to provide the astrophysical community with the first set of data for the SH$^+$--H collisional system. In all the calculations, we have considered all the SH$^+$ energy levels with $N$, $N' \le 10$ and we have included in the calculations all the fundamental rate coefficients with $L \le 12$. The complete set of (de)excitation rate coefficients is available on-line from the LAMDA
\citep{schoier:05} and BASECOL \citep{Dubernet:13} websites.

\subsection{Fine and hyperfine structure excitation}

\begin{figure}
\begin{center}
\includegraphics[width=9cm,angle=0.]{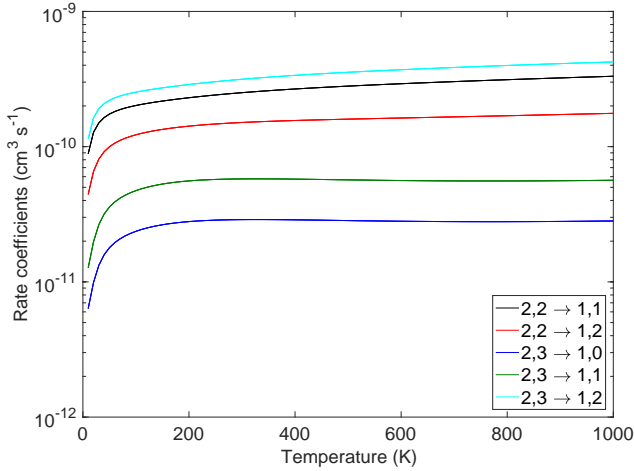}
\caption{Temperature variation of the fine structure resolved de-excitation rate coefficients for the SH$^+$ molecule in collision with H for selected  $N=2,j \to N'=1,j'$ transitions.}
\label{fig2}
\end{center}
\end{figure}

The thermal
dependence of the fine structure resolved state-to-state SH$^+$--H rate coefficients is
illustrated in Fig. \ref{fig2} for selected  $N=2,j \to N'=1,j'$ transitions.

The temperature variation of the de-excitation rate coefficients is relatively smooth except at low temperature ($T < 50 $K) where they increase rapidly. The weak temperature dependence of the rate coefficients (except at low temperature) could have been anticipated, on the basis of Langevin theory for ion--neutral interactions.

A strong propensity rule exists for $\Delta j = \Delta N$ transitions.
Such $\Delta j = \Delta N$ propensity rule was predicted theoretically \citep{alexander:83} and is general for
molecules in the $^{3}\Sigma^{-}$ electronic state. It was also observed previously for the O$_{2}$(X$^3\Sigma^-$)-He
\citep{lique:10}, NH(X$^3\Sigma^-$)--He \citep{Tobola:11} or OH$^+$--H \citep{Lique:16} collisions

\begin{figure}
\begin{center}
\includegraphics[width=9cm,angle=0.]{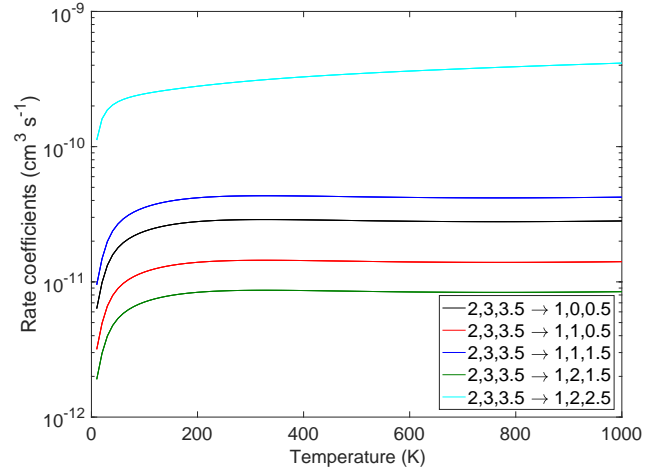}
\caption{Temperature variation of the hyperfine structure resolved de-excitation rate coefficients for the SH$^+$ molecule in collision with H for the $N=2,j=3,F=3.5 \to N'=1,j',F'$ transitions.}
\label{fig3}
\end{center}
\end{figure}

Figure \ref{fig3} presents the temperature variation of the hyperfine structure resolved state-to-state SH$^+$--H rate coefficients
for selected $N=2,j=3,F=3.5 \to N'=1,j',F'$ transitions. 

For $\Delta j = \Delta N$ transitions, we have a strong propensity rule in favor of $\Delta j= \Delta F$ hyperfine transitions . This trend is the usual trend for open-shell molecules \citep{alexander:85,dumouchel:12,kalugina12,Lique:16}. For $\Delta j \ne \Delta N$ transitions, it is much more difficult to find a clear propensity rule. The final distribution seems to be governed by two rules: the rate coefficients show propensity in favor of $\Delta j= \Delta F$ transitions, but are also proportional to the degeneracy ($2F' + 1$) of the final hyperfine level as already found for CN--para-H$_2$ system \citep{kalugina12}. 

\subsection{Comparison with SH$^+$--H$_2$ rate coefficients}

Then, we compare the new SH$^+$--H  rate coefficients with those reported recently for the hperfine excitation of SH$^+$ by H$_2$ \citep{Dagdigian:19}. The SH$^+$ molecule has been observed in media where both atomic and molecular hydrogen are significant colliding partners and this comparison should allow evaluating the impact of the different
collisional partners.  

\begin{figure}
\begin{center}
\includegraphics[width=9cm]{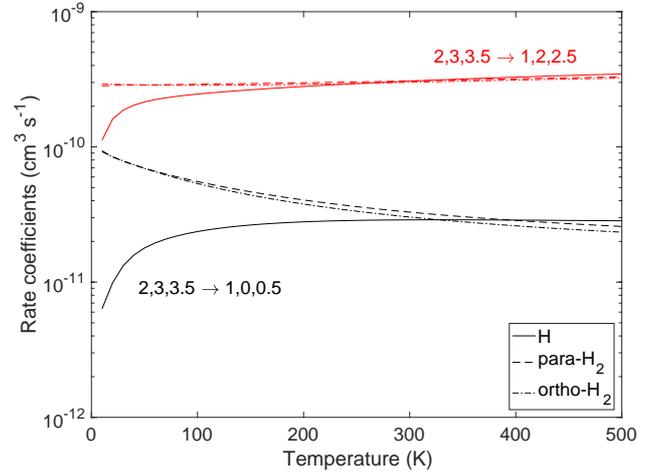}
\caption{Comparison between SH$^+$--H and  SH$^+$--H$_2$ (both para- and ortho-H$_2$) rate coefficients for a selected number of hyperfine ($N=2,j=3,F=3.5 \to N'=1,j',F'$) transitions.}
\label{fig4}
\end{center}
\end{figure}

In Fig. \ref{fig4}, we compare the SH$^+$--H and SH$^+$--H$_2$ (both
para- and ortho-H$_2$) rate coefficients for a selected number of
transitions.

In astrophysical applications, when collisional
data are not available, it is very common to derive collisional data from collisional rate
coefficients calculated for the
same molecule in collision with another colliding partner. Such approach
\citep{lique08b}, consist in assuming that the excitation
cross-sections are similar for both colliding systems and that the rate coefficients differ only by a scaling factor due to the reduced mass which
appears in Eq. \ref{thermal_average}.
Hence,  the following
scaling relationship can be used:
\begin{equation}
k^{{\rm H}} \simeq 1.4 \times k^{{\rm H_2}}
\end{equation}

One can see that, at low temperatures, the rate coefficients for collisions with H do not have the highest magnitude as expected from the scaling relationships. They can even be one order of magnitude weaker.
We also note that the differences between H and H$_2$ rate coefficients depend on the
transitions and on the temperature leading to the impossibility of
extrapolating accurate H collisional data from H$_2$ collisional data, or the reverse. 
Hence, it confirms that it is unrealistic to estimate unknown collisional rate coefficients by simply applying a scaling
factor to existing rate coefficients. This result was previously
observed for water \citep{daniel:15} and ammonia \citep{Bouhafs:17}. 

However, when the temperature increases, the agreement gets better and scaling techniques would lead to a reasonable estimate of the H or H$_2$ rate coefficients for temperatures above 500 K.

\section{Summary and Conclusion}

The fine and hyperfine excitation of SH$^+$ by H have been investigated. We have obtained fine and hyperfine resolved rate coefficients for transitions involving the lowest levels of SH$^+$ for temperatures ranging from 10 to 1000 K. Fine structure resolved rate coefficients present a strong propensity rules in favor of $\Delta j = \Delta N$ transition. The $\Delta j= \Delta F$ propensity rule is observed for the hyperfine transitions.

As a molecule that can be observed from ground-based observatories  \citep{Muller:14}, in the Milky Way and beyond \citep{Muller:17}, we expect that these new data will significantly help in the accurate interpretation of SH$^+$ rotational emission spectra from dense PDRs and massive proto-stars, enable this molecular ion to act as tracer of the energetics of these regions, and of the first steps of the sulfur chemistry.

\begin{acknowledgements}
We acknowledge the French-Spanish collaborative project PICS (Ref. PIC2017FR7). 
F.L. acknowledges financial support from the European Research Council (Consolidator Grant COLLEXISM, grant agree- ment 811363), the Institut Universitaire de France and the Programme National ''Physique et Chimie du Milieu Interstellaire'' (PCMI) of CNRS/INSU with INC/INP co-funded by CEA and CNES.
 The research leading to these results has received funding from
 MICIU under grants No. FIS2017-83473-C2 and AYA2017-85111-P. N.B. acknowledges the computing facilities by TUBITAK-TRUBA. This work was performed using HPC resources from GENCI-CINES (Grant A0070411036).
\end{acknowledgements}
 
\bibliographystyle{aa}

%\bibliography{vanderwaals}

\end{document}